\documentclass[12pt]{article}
\usepackage{times}
\usepackage{geometry}
\usepackage[utf8]{inputenc}
\usepackage{enumitem,amssymb}
\usepackage{ragged2e}
\usepackage{wrapfig}

\usepackage[T1]{fontenc}
\usepackage[usenames,dvipsnames]{xcolor}
\usepackage[breakable, theorems, skins]{tcolorbox}
\tcbset{enhanced}

\DeclareRobustCommand{\mybox}[2][gray!20]{%
\begin{tcolorbox}[   %% Adjust the following parameters at will.
        breakable,
        left=0pt,
        right=0pt,
        top=3pt,
        bottom=2pt,
        colback=#1,
        colframe=#1,
        width=\dimexpr\textwidth\relax, 
        enlarge left by=0mm,
        boxsep=5pt,
        arc=0pt,outer arc=0pt,
        ]
        #2
\end{tcolorbox}
}

\newlist{thematic}{itemize}{8}
\setlist[thematic]{label=$\square$}
\usepackage{pifont}

\usepackage{verbatim}
\usepackage{natbib}
\usepackage{mathtools}
\newcommand{\apj}{ApJ}

\newcommand{\apjs}{ApJS}
\newcommand{\aj}{AJ}

\newcommand{\mnras}{MNRAS}

\newcommand{\araa}{ARA\&A}

\newcommand{\lyaf}{Ly$\alpha$ forest}

\newcommand{\nhi}{$N_{\rm HI}$}

\newcommand{\mlnhi}{\log N_{\rm HI}}

\newcommand{\etal}{\textit{et al.}}

\newcommand{\hst}{{\em HST}}

\newcommand{\hi}{H$\;${\small\rm I}\relax}

\newcommand{\cii}{C$\;${\small\rm II}\relax}
\newcommand{\ciii}{C$\;${\small\rm III}\relax}

\newcommand{\oii}{O$\;${\small\rm II}\relax}
\newcommand{\oiii}{O$\;${\small\rm III}\relax}

\begin{document}
\raggedright
\huge
Astro2020 Science White Paper \linebreak

Following the Metals in the Intergalactic  and Circumgalactic Medium over Cosmic Time
\linebreak
\normalsize

\noindent \textbf{Thematic Areas:} \hspace*{60pt} $\square$ Planetary Systems \hspace*{10pt} $\square$ Star and Planet Formation \hspace*{20pt}\linebreak
$\square$ Formation and Evolution of Compact Objects \hspace*{31pt} $\checkmark$ Cosmology and Fundamental Physics \linebreak
  $\square$  Stars and Stellar Evolution \hspace*{1pt} $\square$ Resolved Stellar Populations and their Environments \hspace*{40pt} \linebreak
  $\checkmark$    Galaxy Evolution   \hspace*{45pt} $\square$             Multi-Messenger Astronomy and Astrophysics \hspace*{65pt} \linebreak
  
\textbf{Principal Author:}

Name: Nicolas Lehner	
 \linebreak						
Institution:  University of Notre Dame
 \linebreak
Email: \texttt{nlehner@nd.edu}
 \linebreak

\textbf{Co-authors:} 
  \linebreak
\noindent
Joseph N. Burchett, Univ. of California - Santa Cruz, \texttt{burchett@ucolick.org}\\
\hangindent=0.7cm
J.\ Christopher Howk, University of Notre Dame, \texttt{howk.1@nd.edu}\\
John M.\ O'Meara, W.\ M.\ Keck Observatory, \texttt{jomeara@keck.hawaii.edu}\\
\hangindent=0.7cm
Molly S. Peeples, Space Telescope Science Institute / Johns Hopkins University, \texttt{molly@stsci.edu} \\
\hangindent=0.7cm
Marc Rafelski, Space Telescope Science Institute / Johns Hopkins University, \texttt{mrafelski@stsci.edu}\\
\hangindent=0.7cm
Joseph Ribaudo, Utica College, \texttt{jsribaud@utica.edu}\\
Sarah Tuttle, University of Washington, Seattle, \texttt{tuttlese@uw.edu}
\linebreak

\justifying
\vspace{-0.3 truecm}
\textbf{Abstract:}
The circumgalactic medium (CGM) of galaxies serves as a record of the influences of outflows and accretion that drive the evolution of galaxies. Feedback from star formation drives outflows that carry mass and metals away from galaxies to the CGM, while infall from the intergalactic medium (IGM) is thought to bring in fresh gas to fuel star formation. Such exchanges of matter between IGM-CGM-galaxies have proven critical to producing galaxy scaling relations in cosmological simulations that match observations. However, the nature of these processes, of the physics that drives outflows and accretion, and their evolution with cosmic time are not fully characterized. One approach to constraining these processes is to characterize the metal enrichment of gas around and beyond galaxies. Measurements of the metallicity distribution functions of CGM/IGM gas over cosmic time provide independent tests of cosmological simulations. We have made great progress  over the last decade as direct result of a very sensitive, high-resolution space-based UV spectrograph and the rise of ground-based spectroscopic archives. We argue the next transformative leap to track CGM/IGM metals during the epoch of galaxy formation and transformation into quiescent galaxies will require 1) a larger space telescope with an even more sensitive high-resolution spectrograph covering both the far- and near-UV (1,000--3,000 \AA); and 2) ground-based archives housing science-ready data.

\thispagestyle{empty}
\pagebreak
\setcounter{page}{1} % max of five pages including figures

\section{Background and Motivation}
A primary challenge in the next decade will be to understand the intimate connection of the intergalactic medium (IGM), the circumgalactic medium (CGM), and galaxies. While we have made significant progress in characterizing each of these components, much work remains to be done to understand how the IGM feeds the CGM and galaxies and vice-versa, and how these processes change across cosmic time. We know, for example, large-scale outflows are ubiquitous at all epochs \citep[e.g.,][]{shapley03,weiner09,steidel10,rubin14}, but we are far from fully understanding the physics of galaxy feedback, its influence on galaxy evolution, and its signatures within the CGM. Similar logic applies to the accretion of gas from the IGM through the CGM: we know it must occur, but the detailed physics is far from constrained.

The distribution of metals with environment provides a way to probe the IGM/CGM/galaxy interconnection. Metals are ``tracer particles'' that can be used to tag the origins of gas in different environments: the metallicity (and relative metal abundances) of components of the CGM/galaxy ecosystem inform us of their origins, differentiating between matter impacted by (or not) the polluting influence of stars. The metal enrichment level of a structure tells us how far metals have traveled from their formation site, how important the influence of the central galaxy compared to the IGM or satellite ejecta is in shaping the structure, and---with enough data---how homogeneously those influences are throughout the CGM of galaxies. Metal rich gas at large distances from galaxies suggest outflows or recycling motions are important on large scales \citep[e.g.,][]{tripp11}, while metal-poor gas found deep in the halos of galaxies trace the influence of fresh IGM accretion \citep[e.g.,][]{berg19,chen19}.

\begin{figure}
    \centering
    \includegraphics[width=\textwidth]{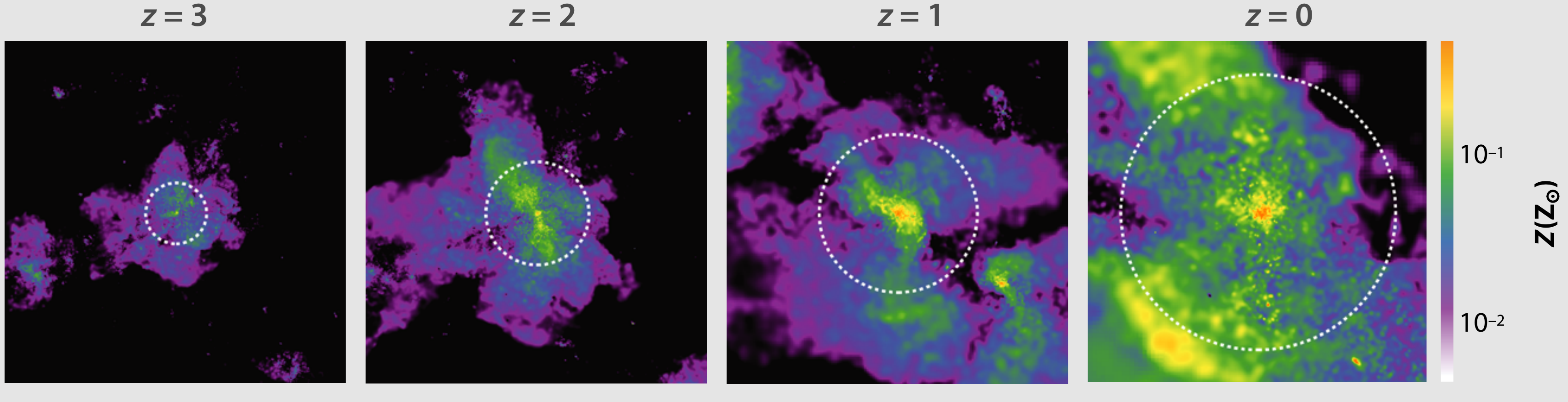}
    \caption{\small Distribution of the metallicity in the CGM of a galaxy at 4 different epochs in the EAGLE simulations \citep{schaye15,oppenheimer16}. The dotted white circle encloses the virial radius of that galaxy at each epoch (each panel is 1 Mpc physical, adapted from Fig.~3 in \citealt*{tumlinson17}). At any redshift, the CGM is filled with low and high metallicity gas, but only at $z\gtrsim 1$ does the IGM start to have metal-enriched gas. These findings can be directly confronted with empirical results. 
    \label{f-eagle}} 
\end{figure}

Fig.~\ref{f-eagle} shows the distribution of the metallicity in a cosmological simulation at different epochs. At all epochs, the gas around galaxies is a mixture of very low metallicity ($\log Z/Z_\odot <-2$) and metal-rich ($\log Z/Z_\odot >-1$) matter. Therefore  empirical studies must be sensitive to both high and low metallicities at all redshift to truly characterize the metals in the diffuse regions of the universe. Empirical mapping the metallicity as shown in Fig.~\ref{f-eagle} is unfortunately unlikely to happen in the next decade (but see Astro2020 White Papers by Tuttle \etal\ and Zaritsky \etal). However, QSO absorption-line techniques provide an efficient way to determine the metallicity of gas over nearly 10 orders of magnitude in \hi\ column density (i.e., from the densest/most-neutral to the most-diffuse/ionized regions of the universe).  So although we cannot (yet) map directly the metallicities of the gas in emission (outside of the central regions of galaxies), we can statistically ``map'' the metallicity distribution with \hi\ column density using QSO absorption-line techniques, providing some information on the metal distribution between galaxies, their CGM, and the IGM \citep{wotta19,lehner19}. Such absorption line measurements can be made over most of the age of the universe, allowing us to determine how metals in these components change over cosmic time \citep[e.g.][]{rafelski12}. 

Our characterization of the IGM/CGM has advanced significantly since the Astro2010 decadal survey over all redshifts (see \S\ref{s-result}). Several complementary factors have allowed us to make such progress, notably: 1) a space-based telescope ({\it Hubble}) with a sensitive UV medium resolution spectrograph (Cosmic Origins Spectrograph---COS); 2) high resolution spectrographs on large ground-based telescopes (Keck/HIRES, VLT/UVES) with {\it public archives}; and 3) availability of high performance computers (estimating metallicities often requires large ionization corrections---see, e.g., \citealt{crighton13,fumagalli16,wotta19}, and high-performance computing has allowed us to apply Bayesian MCMC techniques to make this process much more robust and systematic). 

\mybox[green!20]{
We argue that progress in mapping the IGM/CGM/galaxy connection through metals will require an even larger collecting area telescope with a sensitive high-resolution FUV and NUV spectrograph, {\it and}\ public archives hosting science-ready datasets. Combined with the advent of ground- and space-based integral field and multi-object spectrographs (presented in other Astro2020 White Papers, see, e.g., Peeples \etal, Tumlison \etal,  Tuttle \etal), this will enable a new revolution of our understanding of the chemical histories of galaxies, their CGM, and the IGM before, during, and after the peak of the cosmic star formation and the epoch of galaxy transformation.
}
\vspace{-0.6 truecm}

\section{The Astro2010 Revolution: UV and Spectroscopic Archives}\label{s-result}

Paradoxically, the metallicity of the IGM is better characterized at high than low redshift. At high $z\sim 2$--4, thanks to high-S/N QSO spectra obtained with Keck and VLT, the metallicity of the IGM has been relatively well constrained since the early 2000s with $\log Z/Z_\odot \simeq -3$ at $z \sim 2.5$--3 (e.g., \citealt{ellison00,schaye03,aguirre04,simcoe04}). At $z\lesssim 0.5$, a canonical metallicity of the IGM was often assumed to be $\log Z/Z_\odot \simeq -1$, simply because that was  among the lowest known metallicities determined in QSOs absorbers with $\mlnhi \lesssim 14.5$. That ``canonical'' metallicity has often been used as a plausible metallicity of the warm-hot IGM for the estimation of the baryons in the various gas phases of the universe (e.g., \citealt{shull12}). However, at low \hi\ column densities ($\mlnhi <14.5$, i.e., the IGM), none of the existing UV spectrographs on the {\it Hubble Space Telescope} (\hst) have the sensitivity to estimate metallicities below 10\% solar at $z<1$ (\citealt{lehner18}). 

While the full metallicity range of the IGM at $z\lesssim 2$ remains unknown, surveys have changed our view of how the metals are distributed around low-redshift ($z\lesssim 1$) galaxies. Galaxy-absorber cross correlations tell us that low-\hi\ column density, \nhi, systems predominantly probe the IGM (esp. at $\mlnhi \lesssim 14.5$; e.g., \citealt{chen00, tejos14}), while the higher column density gas probes the CGM of galaxies \citep[see, e.g., Figure 19 from][]{lehner19}. 

This means that it is possible to study different structures of the Universe by targeting absorbers with different \nhi. However, probing the regions that matter most to galaxies---the inner regions with the highest densities---requires a large sample of QSO spectra and a sensitive UV spectrograph. This is because 1) the number of absorbers tracing the dense inner regions of galaxies is lower by orders of magnitude than \lyaf\ lines; and 2) $\mlnhi \gtrsim 16.8$ absorbers can substantially reduce the UV flux in QSO spectra, making it difficult to measure the EUV and FUV lines that provide a disproportionate amount of the information:

\mybox{
{\bf The \hst\ COS revolution}: Characterization of IGM and CGM gas requires EUV/UV diagnostics (e.g., \ciii\ $\lambda$977, \cii\ $\lambda \lambda$1334, 1036, 904, 903, \oii\ $\lambda \lambda$832,833,834, \oiii\ $\lambda \lambda$702, 832, \hi\ Lyman series,...) because the gas is largely ionized, requiring ionization modeling to determine the metallicity of the gas \citep[e.g.,][]{lehner13,fumagalli16,wotta19}.  At $z\lesssim 1.5$, most of these diagnostics, however, remain in the observed UV bandpass, requiring space-based telescopes (see Figure~6 of \citealt*{tumlinson17}). 

COS on \hst\ has transformed our ability to study the CGM at $z\lesssim 1$ for two main reasons: it has tremendous sensitivity (COS is a factor 10 times more sensitive than STIS) at a spectral resolution that is sufficient to enable studies of gas-phase physics. The great sensitivity provides access to a much larger sample of (fainter) QSOs for probing the CGM/IGM, while the spectral resolution is key for deriving column densities and kinematics (not just equivalent widths). The combination of the two has yielded an archive (HSLA, \citealt{peeples17}) of several hundred QSO sightlines---already $\sim$40 times larger than the STIS E140M archive. This archive is far from ideal, as many of the spectra have very low S/N and/or incomplete wavelength coverage. However, it represents the first UV archive large enough for statistical studies of absorber metallicities, with sufficient sensitivity to survey the low-$z$ CGM without metallicity bias.
}

At $z\gtrsim 2$, the UV and some EUV resonance lines are shifted in the optical bandpass. High-resolution spectrographs on large ground-based 6.5--10-m telescopes (Keck, VLT, LBT, Magellan) can be used to study this epoch. Specific surveys using these large facilities have allowed progress to be made (e.g., \citealt{steidel90,prochaska15}), but major results have also been made as a result of the ground-based data being released in science-ready archives: 

\mybox{
{\bf Ground-based archives}: 
Despite their significant impact on high-resolution spectroscopic studies of the IGM and CGM over the last 2$+$ decades, ground-based facilities are still not reaching their full potential: their archives are generally collections of raw data or even non-existent.  Recent years have seen considerable efforts by small teams---notably KODIAQ \citep{lehner14,omeara15,omeara17} and SQUAD \citep{murphy19}---to return high-resolution QSO spectra from Keck/HIRES and VLT/UVES to the community in a science-ready form.  

Despite these efforts, significant amounts of ground-based IGM+CGM data remain unavailable to the community.  Missing from the absorption line studies to date are commensurate samples of science-ready imaging and lower-resolution spectroscopy of the galaxies producing the absorption seen in QSO spectra.  Turnkey, high-volume, high-quality pipelines for multi-object spectrographs (e.g., Keck/LRIS) are only now in development, again by individual teams of dedicated astronomers.  Future archival progress should not rely on small teams alone: observatories should commit resources to ensuring the longevity of their data, both for extant and future instruments. Funding agencies should support these efforts, both by expanding archival data efforts  and by enhancing instrument funds to require archival delivery of {\em science-ready} data.  Even if \textit{no new data} are obtained, the potential impact on the study of the interconnection between the galaxies, IGM, and  CGM by archival data is immense, and should be realized. {\it The MAST motto---``maximizing scientific accessibility and productivity of astronomical data''---should be the standard for all observatories.} 
}

\begin{figure}
    \centering
    \includegraphics[width=\textwidth]{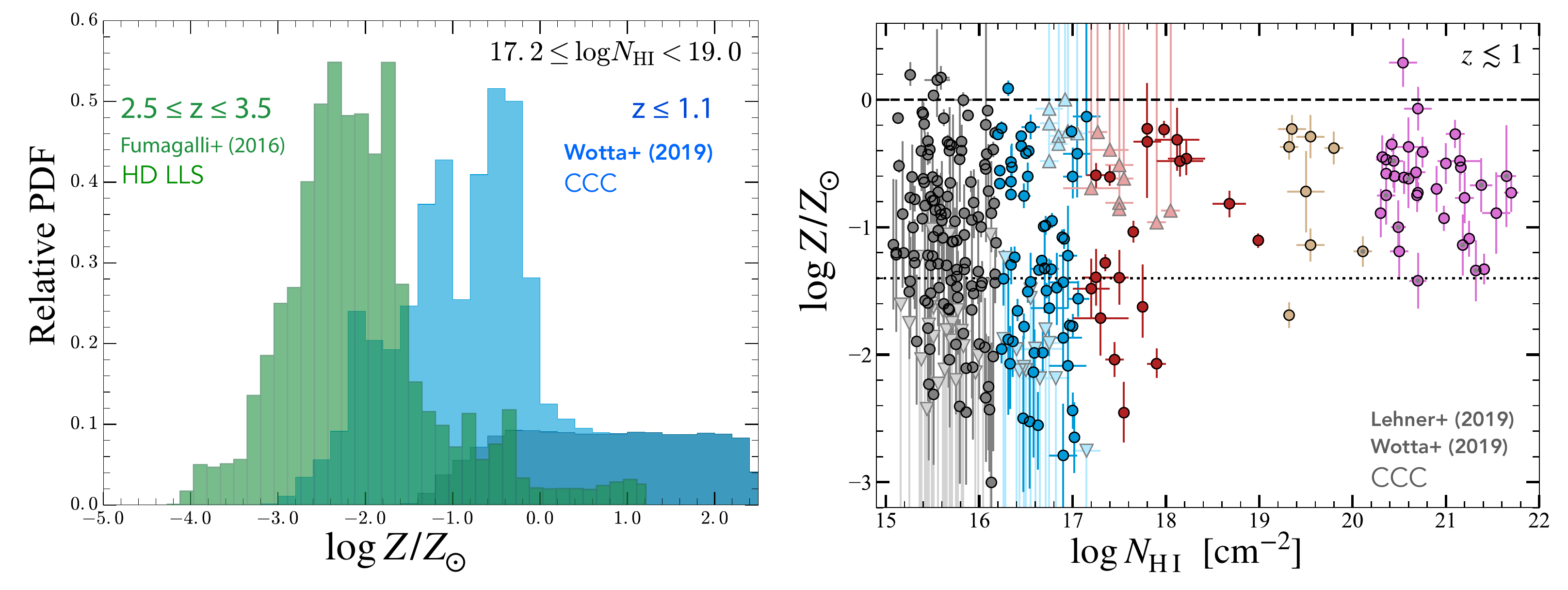}   
    \caption{{\it Left}: Evolution of the metallicity  distributions of the absorbers with $17.2\le \mlnhi \le 18.5$ (from the HD LLS and CCC surveys). {\it Right}: Metallicity evolution versus \nhi\ at $z\lesssim 1$ (from CCC). Prior to the installation of COS on \hst, the only known metallicities at $z\lesssim 1$ were for absorbers with $\mlnhi \gtrsim 19.5$ and $\sim 6$ absorbers with $\mlnhi \lesssim 18$. The IGM regime ($13 \le \mlnhi \le 14.5$) at $z\lesssim 1$ will require to reach S/N\,$\gtrsim$100--200 in high-resolution ($R\sim 30,000$--50,000) UV spectra, only achievable with a large collecting area spaced-based telescope. 
    \label{f-obs}} 
\end{figure}

As an example of the convergence of large UV and ground-based spectral archives,  the sample of absorbers with $\mlnhi \lesssim 18.5$ where we can derive the metallicity has increased by a factor over 40  at $z\lesssim 1$ (\citealt{lehner19}) ($\sim$6 prior to COS to $>$260 now). Fig.~\ref{f-obs} shows the metallicity distribution from surveys of high-density regions of the universe at two different epochs (HD LLS survey: \citealt{prochaska15,fumagalli16}; KODIAQ\,Z survey: \citealt{lehner14,lehner16}; CCC survey: \citealt{lehner18,lehner19,wotta19}; as well as, e.g., \citealt{kanekar14,peroux08,som15} for stronger \hi\ absorbers).

As illustrated in Fig.~\ref{f-obs}, once we have statistically significant  samples in hand, the findings become transformative. These data demonstrate more than half of the absorbers with $\mlnhi \lesssim 18$ have very low metallicities ($\log Z/Z_\odot <-1.4$) that were undetectable in previous work. This demonstrates the CGM of $z\lesssim 1$ galaxies have both low and high metallicities and is thus poorly mixed. Since the metallicity distribution at $2\lesssim z\lesssim 3.5$ peaks around $\log Z/Z_\odot \simeq -2$ \citep{cooper15,glidden16,lehner16,fumagalli16}, this implies some of the CGM gas around galaxies has had little or no net chemical enrichment over $\sim$6 Gyr despite the large-scale outflows from star-forming galaxies, implying that the CGM must be replenished with metal-poor IGM gas. Furthemore, this implies the average metallicity of the IGM at low redshift is likely much lower than previously thought. According to cosmological simulations (see Fig.~\ref{f-eagle}), the metallicity of the IGM at $z\sim 0.5$ peaks around $\log Z/Z_\odot \simeq -2.3$ with a very large spread from $\log Z/Z_\odot <-3$ to  $\log Z/Z_\odot>-0.5$ (\citealt{lehner19}).

With large samples of absorbers, we can finally address how the metals are distributed in the cosmic structure. This directly provides new ways to confront the results from cosmological simulations as the one shown in Fig.~\ref{f-eagle}. The strong feedback required to explain the properties in the disks of galaxies in current cosmological simulations seems to cause an overabundance of metal-enriched gas in their CGM \citep{lehner19,wotta19}. Recent very high-resolution zoom simulations also show some quantitative changes, which can directly impact  the mixing of metals, especially in the cooler and denser regions of the CGM (e.g., \citealt{corlies19,hummels19,peeples19,suresh19,vandevoort19}), further highlighting the need to understand the distribution of metals in the CGM and IGM at all epochs in both cosmological simulations and the observable universe. We have entered an era where cosmological simulations can be confronted not only by the optical properties of the galaxies but also the gas properties in their CGM and in the IGM.

\vspace{-0.4 truecm}
\section{The Future: A Complete Census of Metals Over Cosmic Time}

\mybox[red!20]{Future metallicity censuses will focus on IGM/CGM metals over $0 \lesssim z\lesssim 2$ in order:

$\bullet$ to understand the mixture of metallicities in the CGM during the peak of cosmic star formation and the era of galaxy transformation;\\
$\bullet$ to trace the build-up of CGM metals to today; \\
$\bullet$ to assess the unexplored metallicity of the IGM---gas beyond the halos of galaxies.
}

Without a future high-resolution FUV/NUV spectrograph on a large space telescope, none of the above issues can be explored. These get at the driving of star formation through accretion during the peak of cosmic star formation and its mitigation by feedback, the role of the CGM in quenching star formation during the era of galaxy transformation, and the extent to which galaxies pollute the universe with metals beyond their own halos. 

A high-resolution UV spectrograph (covering 1,000--3,000 \AA) on a large LUVOIR-like telescope would provide not only critical insights into galaxies over these redshifts, but transform our understanding of the CGM with which their evolution is coupled. High resolution is critical for deriving physical properties and metallicities; wavelength coverage for access of key EUV/UV ions from now through the peak of cosmic star formation; of aperture for access to a broad range of galaxy types, multiple sight lines through individual galaxies, and high-S/N observations to probe extra-halo (IGM) metals. LUVOIR-A type apertures are sufficient to allow the use of background galaxies to sample the CGM and IGM at unprecedented spatial scales.

The archives of QSO spectra from current and future large optical observatories will allow us to extend these advances to the universe at $z\gtrsim 2$ with larger samples of absorbers. The current archives \citep{omeara15,omeara17,murphy19} can be used to build on existing high-redshift surveys (e.g., \citealt{lehner16,fumagalli16,cooper15,glidden16}). The next few years will see a much expanded census of absorber metallicities at $z\simeq 2.3$--4. However, without a new generation space-based UV telescope, about 75\% of cosmic time will remain poorly characterized, at a critical time where galaxies are transforming and in a redshift interval where the next generation of integral field units will yield a groundbreaking wealth of data on the galaxies (see, e.g., white paper by Peeples \etal).

\pagebreak

\end{document}